\documentclass[prl,aps,preprint,epsfig]{revtex4}
\usepackage{graphicx}
\usepackage{multirow}
\usepackage{threeparttable}
\usepackage{dcolumn}
\usepackage{bm}
\makeatletter

\newcommand{\Rmnum}[1]{\expandafter\@slowromancap\romannumeral #1@}
\makeatother
\begin{document}
\title {Direct observation of valley-hybridization and universal symmetry of graphene with mesoscopic conductance fluctuations}
\author{Atindra Nath Pal, Vidya Kochat, and Arindam Ghosh} \vspace{1.5cm}
\address{Department of Physics, Indian Institute of Science, Bangalore 560 012, India}

\begin{abstract}
In graphene, the valleys represent spin-like quantities and can act as a physical resource in valley-based
electronics to novel quantum computation schemes. Here we demonstrate a direct route to tune and
read the valley quantum states of disordered graphene by measuring the
mesoscopic conductance fluctuations. We show that the conductance fluctuations
in graphene at low temperatures are reduced by a factor of four when
valley triplet states are gapped in the presence of short range potential
scatterers at high carrier densities. We also show that this implies a gate
tunable universal symmetry class which outlines a fundamental
feature arising from graphene's unique crystal structure.

\end{abstract}


\maketitle

%

Quantum interference of electrons (or holes) causes the electrical conductance
$G$ of a disordered metal to fluctuate aperiodically yet reproducibly with
Fermi energy, magnetic field or disorder
configuration~\cite{Birdge_PRL,Feng_Lee_stone,Altshuler_JETP_RMT}. When the sample size is
smaller than the phase coherence length ($L_\phi$), the conductance fluctuates
with a universal magnitude $\sim e^2/h$, the quantum of conductance,
irrespective of material properties, device geometry or dimensionality. In
conventional mesoscopic conductors, such as thin metal films~\cite{Birdge_PRL} or weakly localized semiconductor devices~\cite{arindam3}, both universal conductance fluctuations (UCF) and weak localization effects are
well-understood on the common platform of coherent backscattering of quasiparticle wave functions~\cite{Feng_Lee_stone}.

The scenario is considerably more complex in graphene due to the existence of
two degenerate valleys~\cite{Efetov_UCF}. The hexagonal lattice structure of
graphene contains two basis atoms in its unit cell, which makes the conduction
and valence bands of graphene cross the Fermi level at two inequivalent $K$ and
$K'$ points on the opposite corners of the hexagonal Brillouin zone, leading to
two valleys~\cite{Wallec,Castreneto_RMP}. Consequently, quantum correction to the conductivity in graphene is
determined not by the inelastic processes alone, but also by elastic collision
events that involve the intervalley and intravalley
scattering~\cite{sdsarma_RMP,savchenko_WL,savchenko_antilocalization}.

\begin{table*}[!htb]
\caption{\label{TabA1}Details of the single layer graphene devices}
\label{Tab6.1}
\centering
\begin{threeparttable}[b]
\begin{tabular}{c| c c c c}\hline
 Device & Device area (L$\times$W)\tnote{1}&~ Dirac point\tnote{2} &~electron mobility\tnote{3}&~hole mobility\tnote{3}\\ \hline
 &\\
 Dev~I & $ 2.1\times 2.9$ &-5.5& 12,000& 10,000  \\
 Dev~II & $ 2.5\times 3.6$ &-8& 4,500& 5,700  \\
 Dev~III & $ 3.3\times 5.5$ &40& 2,300&2,800  \\
 &\\
 \hline
\end{tabular}
\begin{tablenotes}
    \item[1] both dimensions in $\mu$m
    \item[2] in Volt
    \item[3] in cm$^2$/Vs at 4~K
\end{tablenotes}
\end{threeparttable}
\end{table*}

Although the signature of UCF in low temperature electrical transport in
mesoscale graphene has been reported in numerous
experiments~\cite{UCF_graphene_solid,UCF_graphene_JPC,UCF_graphene_nanotech,UCF_graphene_acsnano},
a quantitative influence of valleys on UCF has never been observed. Numerical
investigations suggest that the UCF magnitude in graphene should be stronger
than in normal metal, raising doubts whether its magnitude is ``universal'' at
all~\cite{UCF_Beenakker_EPL}. Moreover, UCF is sensitive to the universal
symmetry class of a disordered system which in case of graphene is determined
by time reversal symmetry (TRS) as well as valley
degeneracy~\cite{sdsarma_RMP,Ando_PRL_2002,valley_contrasting_PRL}. Thus UCF may reveal, for example,
whether graphene belongs to the symplectic symmetry class in the absence of
intervalley scattering, or whether time reversal symmetry is broken by ripples
due to substrate roughness~\cite{TRS_breaking_PRL}, or edge magnetism~\cite{Magnetism_shenoy} \textit{etc.}.

In this letter we present the first direct manifestation of the valley coherent states on UCF in monolayer disordered graphene. The key result
is the suppression of UCF magnitude within individual phase coherent boxes by
a factor of four as the carrier density is increased from close to
the Dirac point, where both valley singlet and triplets contribute to the
fluctuations, to the high electron or hole density regime, where short range
potential fluctuations allow only valley singlets to survive. We find the
factor of four suppression to be universal, irrespective of the geometry,
carrier mobility or temperature, indicating it to be a unique and fundamental
property of mesoscopic graphene.

Graphene devices were prepared by standard micro-mechanical exfoliation of
natural graphite on $300$~nm SiO$_2$/Si$^{++}$ wafer surface (see the SEM image of a typical graphene device in the inset of Fig.~1a and our earlier works~\cite{Atin_PRL} for more detail on device fabrication). Here we present
detailed studies on three monolayer graphene devices with varying disorder (see
Table~I for details). The conventional magnetotransport was studied first,
where we used the method described in Ref.~\cite{savchenko_WL} to calculate the
average magnetoconductance (MC) within small gate voltage ($V_{BG}$) windows of
4~V for each transverse magnetic field ($B$). In Fig.~1b we show the quantum
correction to conductivity $\Delta\sigma(B) = \sigma(B) - \sigma(0)$, as a
function of $B$ in Dev~I at three values of $V_{BG}$ which are identified on
the resistance, $R-V_{BG}$ trace for the same device in Fig.~1a. Measurements were performed in two terminal configuration as appropriate for UCF~\cite{Lee_stone_Fuku}, although, for fitting the $B-$dependence of MC we have used the four-terminal resistivity to eliminate the contact resistance. With decreasing
density (moving from region 3 to region 1) MC changes its sign from
positive to negative similar to the observations in
Ref.~\cite{savchenko_antilocalization}. This transition from weak localization
to weak anti-localization indicates that the scattering associated
with short-range defects is stronger at high carrier density and one can tune
the relative strength of the elastic scattering just by changing the carrier
density or gate voltage. On fitting the traces with the theoretical expression~\cite{Mccann_WL}, we find that
$L_\phi$ saturates to a value that is in the order of the device dimension in
all cases below $\sim 4$~K, which is about $5-10$ times longer than the elastic
mean free path~\cite{SOM}.

In order to evaluate the magnitude of conductance fluctuations, we chose
successive gate voltage windows (of equal width $\Delta V_{BG} = 4$~V) within which the average conductance does not vary appreciably, but we have
significant fluctuations for statistically meaningful analysis (up to $\sim
800$ realizations). The conductance variation within a typical window is shown
in Fig.~1c, where the random yet reproducible fluctuations in $G$ has an
amplitude of $\sim e^2/h$, which is the hallmark of UCF. The fluctuations
weaken with increasing temperature ($T$), as illustrated in Fig.~1d. Below $\sim 300$~mK,
$\langle\delta G^2\rangle$ becomes nearly constant as $L_\phi$ itself saturates
to $\sim$ device dimension~\cite{SOM}.

In order to establish the origin of the conductance fluctuations, we measured
the fluctuation magnitude in a small transverse magnetic field. In Fig.~2 the
variation in the normalized noise magnitude $N(B) =
\langle\delta G(B)^2\rangle/\langle G\rangle^2$ with $B$ is shown for Dev~III
at three values of $V_{BG}$ extending from the Dirac region to high carrier
densities (see inset). In all cases $N(B)$ decreases by $\sim$ factor of {\it two},
as $B$ exceeds $\sim 10$~mT which is also the characteristic field scale
associated with the quantum correction to conductivity ($B_\phi = h/eL_\phi^2 \approx 1-4$~mT for Dev~III). In the diagrammatic representation of quantum transport, this can be readily explained by the suppression of the
Cooperon (particle-particle channel) contribution to UCF for $B \gg B_\phi$, while the diffuson (particle-hole channel)
contribution, which is equal in magnitude, remains
unaffected~\cite{Feng_Lee_stone,stone_PRB}. This result has two important
implications: First, in the framework of the random matrix theory, the factor
of two reduction in UCF can be understood by a $B$-induced lifting of TRS that
drives the system from orthogonal or symplectic symmetry class to the unitary
class~\cite{Altshuler_JETP_RMT}. Similar observation here
indicates that at $B = 0$ there is no spontaneous breaking of TRS in our
graphene devices. Second, we conclude that the observed $V_{BG}$-dependent
fluctuation in $G$ is entirely due to UCF, and hence the measured magnitude of
$\langle\delta G^2\rangle$ can be directly used to probe the valley effects and
symmetry class of graphene.

Since valley degeneracy is connected to the carrier density through the nature
of potential scattering~\cite{sdsarma_RMP,Mccann_WL,Efetov_UCF}, we
subsequently measured $\langle\delta G^2\rangle$ over a wide range of carrier
density in all three devices at $T = 10$~mK and $B = 0$. Fig.~3a shows the variation of $\langle\delta G^2\rangle$ with $V_{BG}$ for Dev~I, indicating a weak increase on both sides of the Dirac point. The variation of $\langle\delta G^2\rangle$ with $V_{BG}$ was found to be highly device specific, with opposing trends observed in different devices~\cite{SOM}. The measured $\langle\delta
G^2\rangle$ however arises from the entire sample, and in order to estimate the
UCF magnitude $\langle\delta G_\phi^2\rangle$ within a single phase coherent
box ($L_\phi\times L_\phi$), we have employed the theorem of classical
superposition~\cite{Feng_Lee_stone}, $\langle \delta G^2\rangle/\langle G\rangle^2=(1/N)\times \langle \delta G_{\phi}^2\rangle/\langle G_{\phi}\rangle^2$, where $N = LW/L_{\phi}^2$ is the number of phase coherent boxes in a device of length $L$ and width $W$. Using $\langle G_\phi\rangle = \sigma$, the conductivity and $\langle G \rangle=L\sigma/W$, we get,
\begin{equation}
\label{eq1} \langle \delta G_\phi^2\rangle = \frac{L^3}{W}\times \frac{\langle \delta G^2\rangle}{L_\phi^2}.
\end{equation}

\noindent In Eq.~\ref{eq1}, both $\langle \delta G^2\rangle$ and $L_{\phi}$ are $V_{BG}-$dependent quantities, and we evaluated them experimentally to extract the $V_{BG}-$dependence of $\langle \delta G_{\phi}^2\rangle$. Fig.~3b shows the dependence of $L_{\phi}$ with $V_{BG}$ in Dev~I over the same range of $V_{BG}$. $L_{\phi}$ increases with increasing density, and near the Dirac region it shows a minimum. This is typical for graphene and could be connected to dominance of electron-electron scattering with increasing densities~\cite{inelastic_PRB}. In Fig.~3c we have
shown the variation of $\langle \delta G_\phi^2\rangle$ with $V_{BG}$ in Dev~I at $B = 0$, using experimentally measured $\langle\delta G^2\rangle$ and $L_\phi$ (from MC) at every gate voltage. While the absolute
magnitude of $\langle \delta G_\phi^2\rangle$ is expectedly of the order of
$(e^2/h)^2$ in all cases, the key aspect is the suppression of
$\langle \delta G_\phi^2\rangle$ by a factor of $\approx 4$ as $V_{BG}$ is
varied from the Dirac point towards both high electron or hole density regime.
The characteristic scale of such a suppression seems to follow the crossover of
linear (Coulomb scattering) to sub-linear (short-range
scattering)~\cite{sdsarma_PRL} density-dependence of conductivity (see Fig.~3c and 3d, and vertical guidelines). Similar reductions were observed in the other two devices (Dev~II and Dev~III) as well, shown in Fig.~4a-d. The suppression of noise is often asymmetric
about the Dirac point, probably connected to the difference in mobility and
disorder between the electron and hole side.

\begin{table*}[!htb]
\caption{\label{table1}Values of the symmetry parameters k, s, $\beta$ for
different universality class in case of graphene.}

\begin{tabular*}{1\textwidth}{@{\extracolsep{\fill}} |c|c|c|c|c|c|c|c|c| }
  \hline
& {\multirow{2}{*}{Density}}& \multicolumn{3}{c|}{s}&k&$\beta$&$\frac{ks^2}{\beta}$&Universality class \\\cline{3-5}

& & valley-isospin&Kramer's degeneracy& real spin& & & & \\
\hline
{\multirow{2}{*}{B = 0~T}}
&$n\rightarrow$ low & 2 & 2 & 2 & 1 & 4&16& GSE \\
&$n\rightarrow$ high & 1 & 1 & 2 & 1 & 1&4& GOE \\
\hline
{\multirow{2}{*}{B = 100~mT}}
&$n\rightarrow$ low & 2 & 1 & 2 & 1 & 2&8& GUE\\
&$n\rightarrow$ high & 1 & 1 & 2 & 1 & 2 &2& GUE\\
\hline
\end{tabular*}
\end{table*}

The factor of four suppression in $\langle \delta G_\phi^2\rangle$,
irrespective of mobility, geometry and other device specific details, suggests
a fundamental property of disordered graphene. In a recent
analytical approach, removal of valley degeneracy has been shown to cause an
exact factor of four reduction in the mesoscopic fluctuations in
graphene~\cite{Efetov_UCF}. The underlying physical mechanism can be
schematically presented as in Fig.~4e. At lower density, the scattering is
dominated by the long range Coulomb scattering where the effective
valley-isospin rotational symmetry (SRS) is preserved. Consequently, the
mesoscopic fluctuations receive equal contribution from each of the singlet and
(three) triplet channels of diffusons and Cooperons. However, at higher
densities the intervalley scattering dominates, and the effective SRS is
lifted, and only the singlet diffuson and Cooperon channels contribute. In the
latter case the graphene essentially behaves as a conventional 2D disordered
metal.

The existence of valleys is also expected to result in a nontrivial symmetry property of graphene Hamiltonian that has never been probed in a direct manner. At zero magnetic field the presence of valley isospin rotational symmetry (no intervalley scattering) makes graphene belong to the Gaussian symplectic ensemble (GSE), characterized by a Wigner-Dyson parameter $\beta = 4$, and doubly degenerate isospin and effective time reversed states (Kramer's degeneracy). Removal of valley degeneracy results in a Gaussian orthogonal ensemble (GOE) with $\beta =1$ as well as suppression of effective Kramer's degeneracy~\cite{Ando_PRL_2002}. For open mesoscopic systems, the random matrix theory connects the magnitude of UCF to its symmetry class simply as~\cite{Altshuler_JETP_RMT},

\begin{equation}
\label{eq2}
\langle \delta G_\phi^2\rangle \approx \left (\frac{e^2}{h}\right)^2\frac{ks^2}{\beta}
\end{equation}

\noindent where $k$ is the number of independent eigenvalue sequences of the transmission matrix or Hamiltonian, $s$ is the eigenvalue degeneracy. As shown in the first row of Table~II, factor of four suppression of UCF with increasing carrier density can be explained from the change in symmetry properties of graphene as the valley degeneracy is removed.

Finally, in order to establish the consistency of Eq.~\ref{eq2}, we have measured the $V_{BG}$ dependence of $\langle \delta G_\phi^2\rangle$ at a finite $B$ ($\gg \frac{h}{eL_{\phi}^2}$), which removes the time reversal symmetry and puts graphene in the Gaussian unitary ensemble (GUE, $\beta = 2$) at all densities. As shown for Dev~III at different temperatures (Fig.~5), $\langle \delta G_\phi^2\rangle$ is again suppressed by a factor of four at high densities irrespective of $T$, which can be understood either as gapping of triplet diffusons, or alternatively as, suppression of valley isospin degeneracy in Eq.~\ref{eq2} (see lower row of Table~II).

In conclusion, we have demonstrated for the first time that the mesoscopic conductance
fluctuations in single layer graphene are dependent on the valley hybridized states and physical symmetries
of Hamiltonian. We have shown that the variance of universal conductance fluctuations can be a sensitive probe to this, which
in a single phase coherent box increases by four times near the Dirac region as compared to high density. This sensitivity could be exploited in read-out schemes involving valley states in graphene.

\textbf{Acknowledgement:} We thank Carlo Beenakker and G. Baskaran for useful discussions. We acknowledge the Department of Science and
Technology (DST) for a funded project. ANP and VK thank CSIR for financial
support.


\begin{figure*}[!t]
\begin{center}
\includegraphics [width=0.9\linewidth]{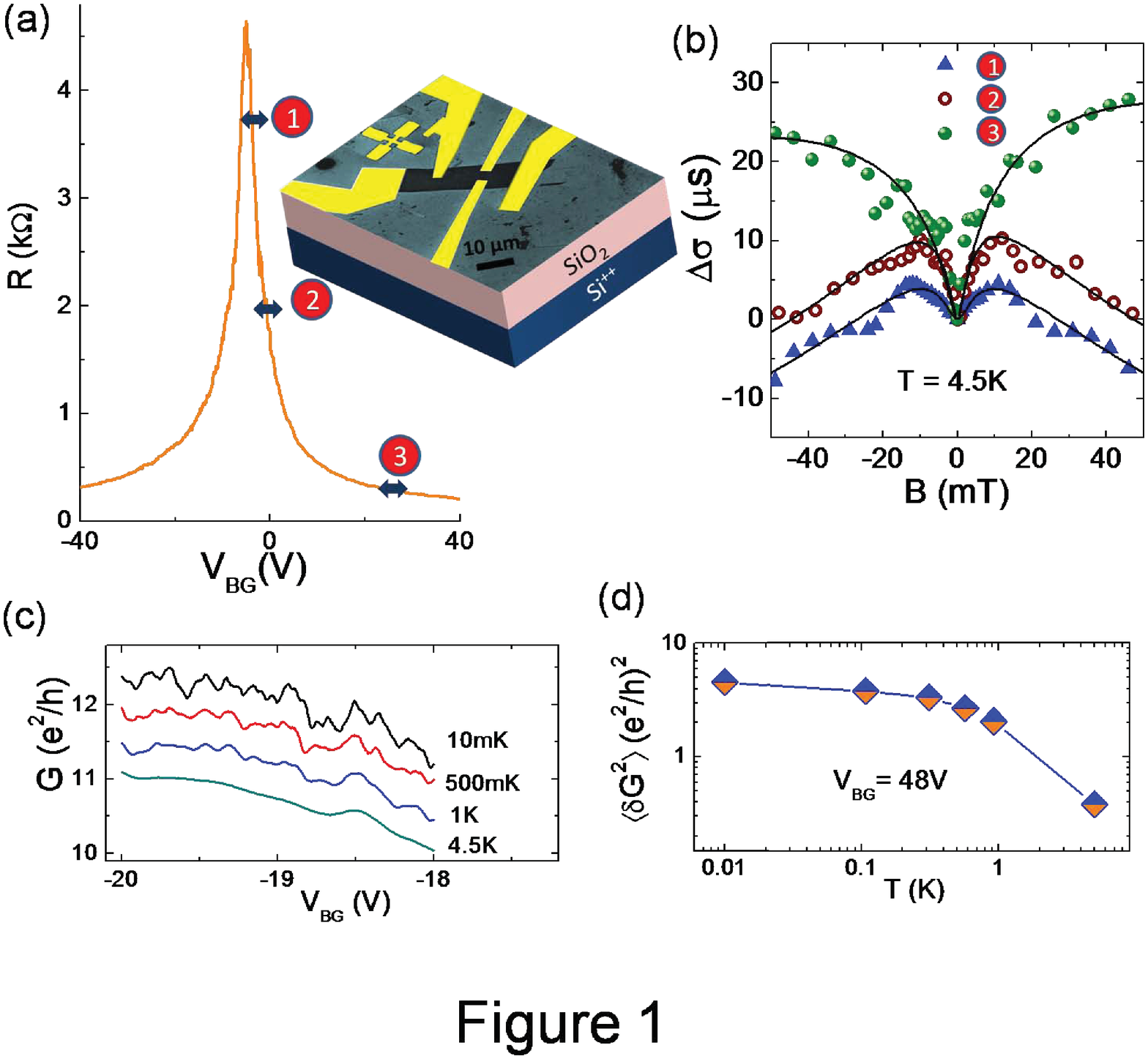}
\end{center}
\caption{ (color online). (a) Resistance ($R$) vs. gate voltage ($V_{BG}$)
characteristics for a single layer device (Dev~I) at $T=10$~mK. Inset shows the schematic along with the false color SEM image of a typical graphene transistor. (b) Evolution of Magnetoconductance with increasing electron density measured at $T = 4.5$~K.
Solid lines are fit to the weak localization theory~\cite{Mccann_WL,SOM}. (c) Conductance fluctuations as a function of gate voltage
for Dev~I, shown for various temperatures. Curves at different temperatures are
shifted for clarity. (d) Variance in conductance, calculated from the
conductance fluctuations in gate voltage range $-50$~V to $-46$~V for Dev~I,
plotted as a function of temperature. } \label{fig1}
\end{figure*}

\begin{figure*}[!htb]
\begin{center}
\includegraphics [width=0.9\linewidth]{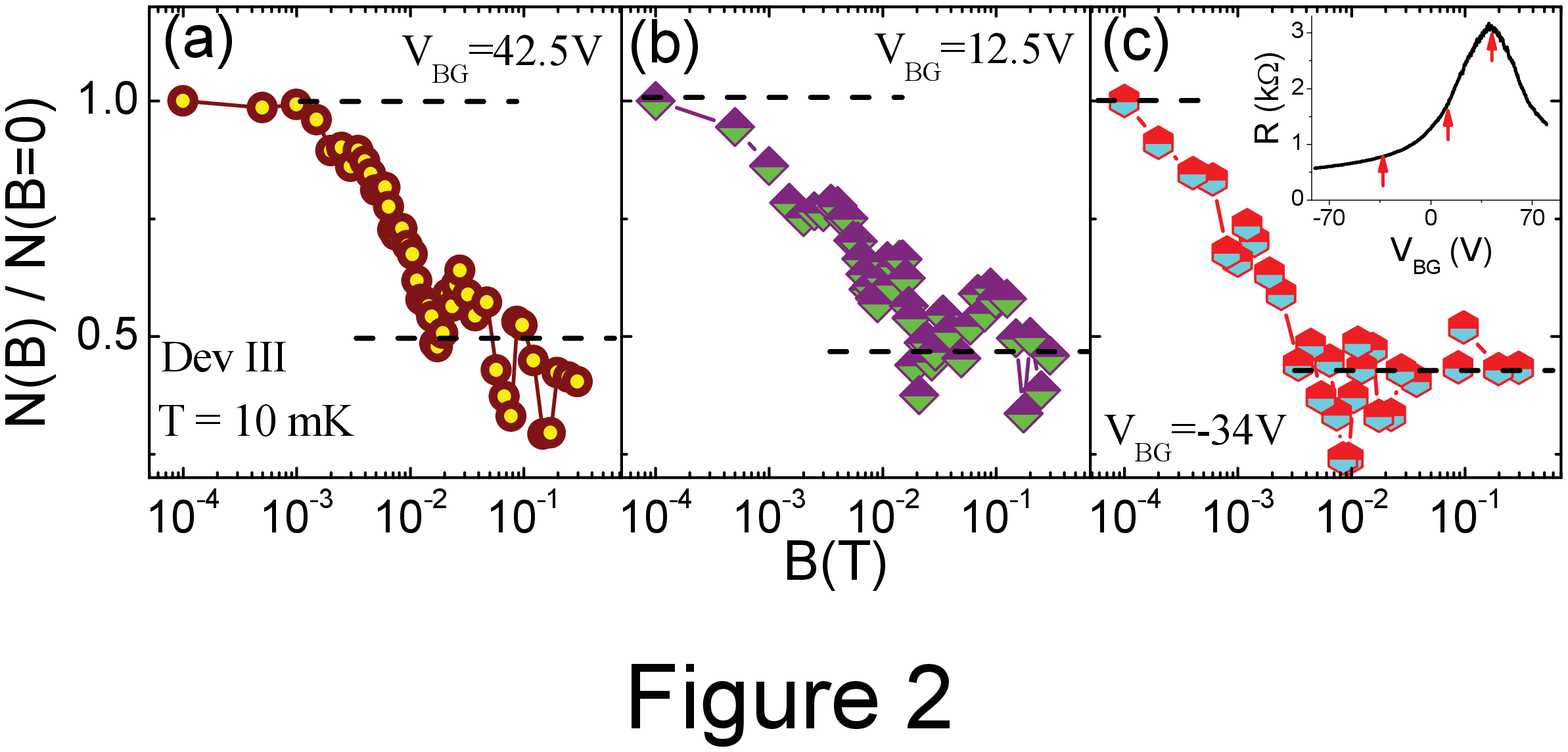}
\end{center}
\caption{ (color online). Magnetic field dependence of the magnitude of conductance fluctuations for Dev~III at $T = 10$~mK at
various gate voltages: (a) 42.5~V, (b) 12.5~V, and (c) -34~V. For comparison, we have plotted
the ratio $N(B)/N(B=0)$, with $N(B)=\langle\delta G^2\rangle/\langle G\rangle^2$ at the magnetic field $B$. Inset shows the $R-V_{BG}$ characteristics of Dev~III at $T = 10$~mK with arrows indicating the three measured regions. } \label{fig2}
\end{figure*}

\begin{figure}[!t]
\begin{center}
\includegraphics [width=0.5\linewidth]{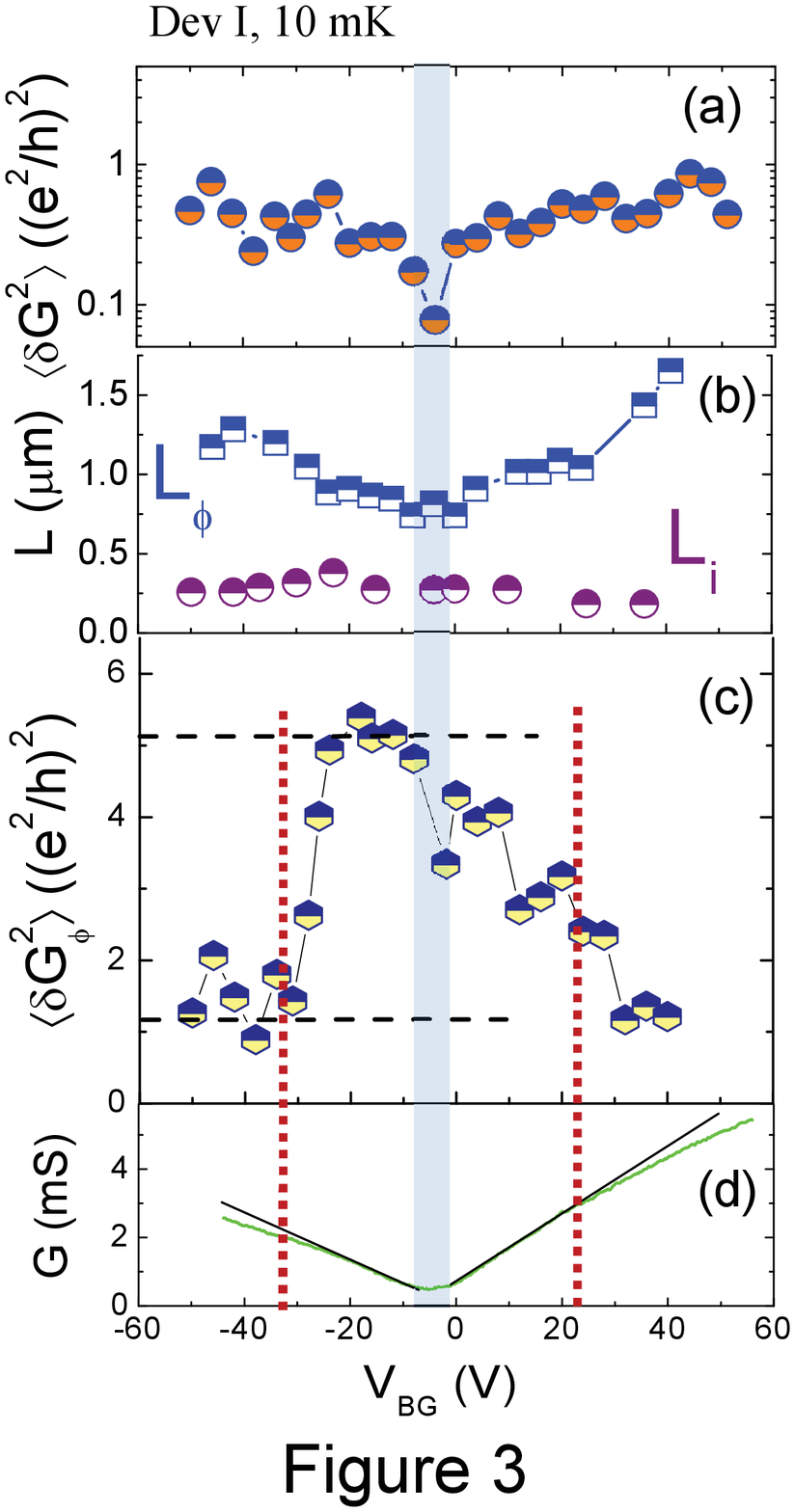}
\end{center}
\caption{(color online). (a) Variance in conductance ($\langle \delta G^2 \rangle$) vs. gate voltage ($V_{BG}$) at
10~mK for Dev~I. (b) Gate voltage dependence of $L_{\phi}$ and $L_i$ at $T = 10$~mK, extracted from the weak localization fits, are shown for Dev~I. (c) The variance in conductance in a phase coherent box of area $L_{\phi}^2$, $\langle \delta G_{\phi}^2 \rangle$, extracted from (a) and (b), is plotted with gate voltage for Dev~I. The factor of four is
indicated by the dashed lines. (d) Conductance ($G$) vs. gate voltage ($V_{BG}$) for the device at $T = 10$~mK. The solid line indicates
the linear region in both electron and hole sides. The vertical dotted lines in (c) and (d) indicate the densities
where the short-range scattering dominate. The shaded region indicates the inhomogeneous region near the Dirac point.  } \label{fig3}
\end{figure}

\begin{figure*}[htb!]
\begin{center}
\includegraphics [width=0.9\linewidth]{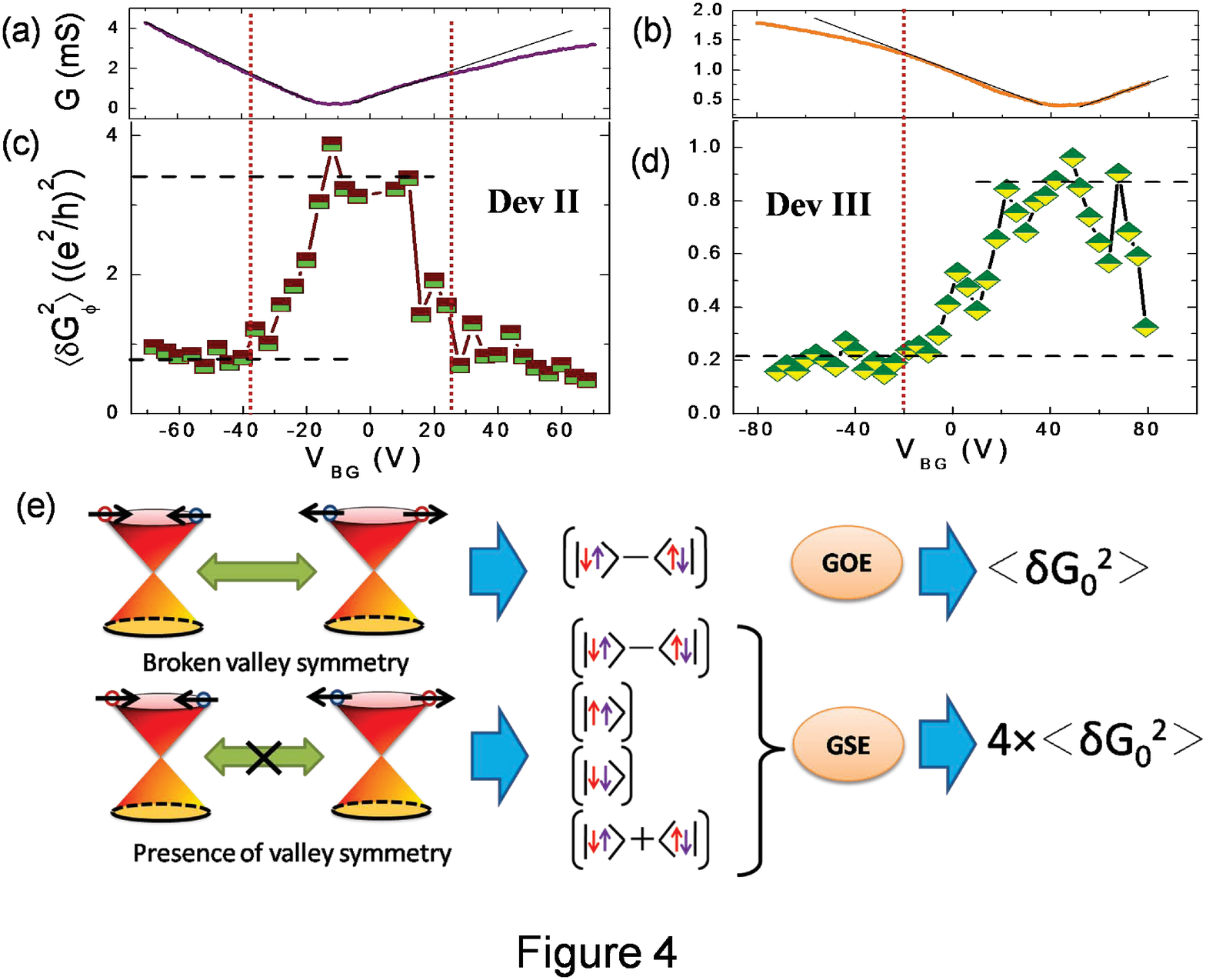}
\end{center}
\caption{(color online). (a)-(b) Conductance ($G$) vs. gate voltage ($V_{BG}$) at
10~mK for Dev~II and Dev~III, respectively. The solid line indicates
the linear region in both electron and hole sides. (c)-(d) The variance
from conductance fluctuations in a phase coherent box of area $L_{\phi}^2$
is plotted with gate voltage for Dev~II and III, see details in Table~I. The factor of four is
indicated by the dashed lines. The vertical dotted lines indicate the densities
beyond which the short-range scattering dominates. (e) Schematic describing the crossover from orthogonal to
symplectic universality class depending on the presence of short range
scattering which breaks the effective valley-isospin rotational symmetry with $\langle \delta G_0^2\rangle$ being the variance in conductance for a single phase coherent box at very high density.}
\label{fig4}
\end{figure*}

\begin{figure*}[htb!]
\begin{center}
\includegraphics [width=0.9\linewidth]{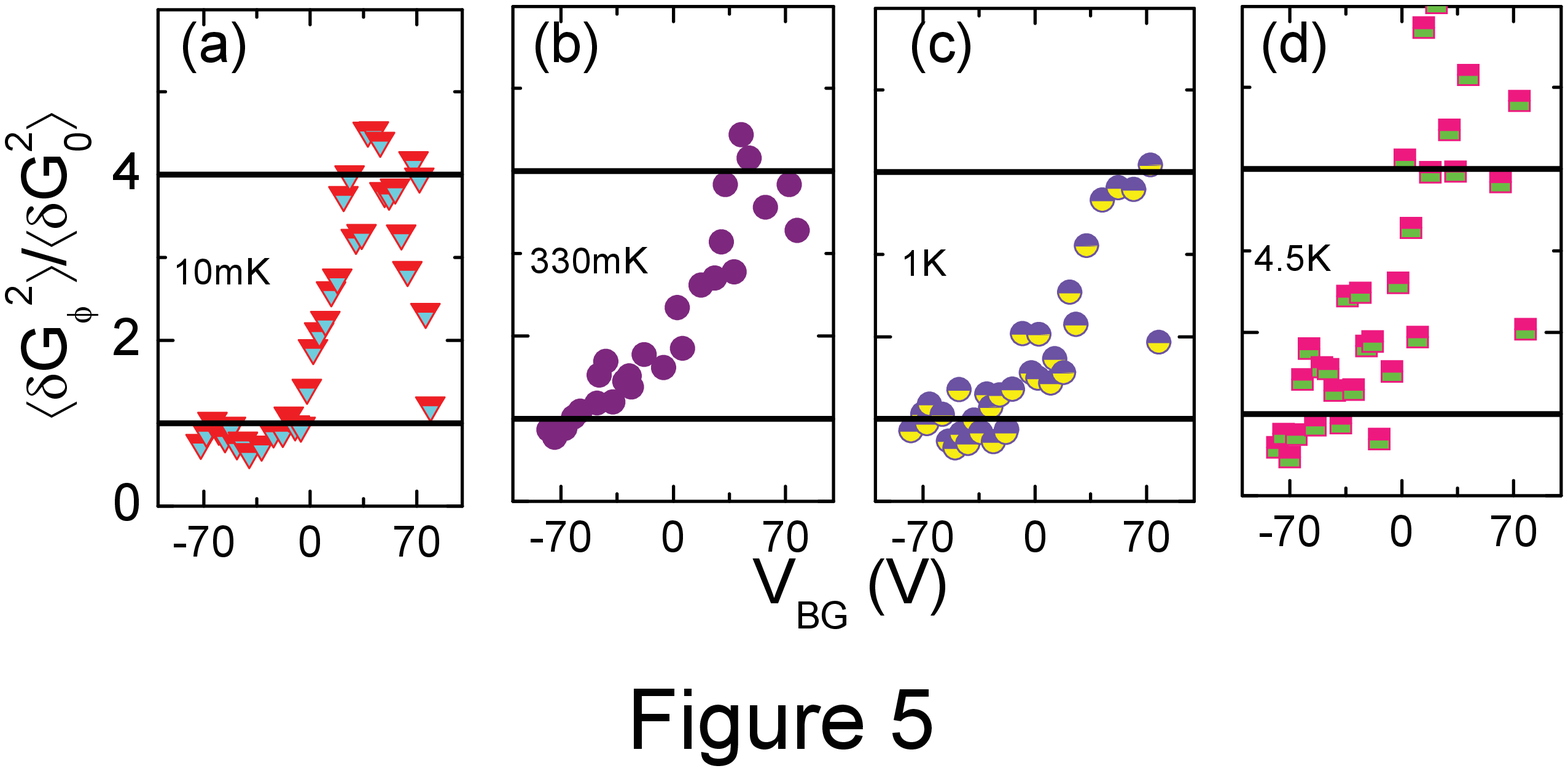}
\end{center}
\caption{(color online). (a)-(d) Gate voltage dependence of the ratio $\langle\delta G_{\phi}^2\rangle/\langle\delta G_0^2\rangle$ for Dev~III,
plotted for various temperatures at $B = 100$~mT, where $\langle \delta G_0^2\rangle$ is the variance in the conductance for a single phase coherent box at $V_{BG}=-68$~V. }
\label{figure2}
\end{figure*}

\end{document}


\title {\huge{Supplementary Information}\\\vspace{5cm}
Direct observation of valley-hybridization and universal symmetry of graphene with mesoscopic conductance fluctuations\\
}
\author{Atindra Nath Pal, Vidya Kochat, and Arindam Ghosh\\
Department of Physics, Indian Institute of Science,\\
 Bangalore 560 012, India. \\
 }

\maketitle{}


\newpage

\section{Magnetoconductivity measurements}\label{sec.1}

For magnetoconductivity (MC) measurements we have averaged out the fluctuations over a gate voltage window of $\Delta V_{BG}=4$~V, following Ref.~\cite{savchenko_antilocalization}, to get the conductivity correction $\Delta \sigma=\langle\sigma (V_{BG},B)-\sigma (V_{BG},0)\rangle_{\Delta V_{BG}}$. The quantum correction depends not only on the dephasing rate $\tau_\phi^{-1}$, but also intervalley ($\tau_i^{-1}$) and intravalley scattering rate ($\tau_*^{-1}$). To fit the MC data we have used the weak localization theory for graphene, given in Ref.~\cite{Mccann_WL},
\begin{equation}
\label{eq1} \Delta \sigma (B)=\frac{e^2}{\pi h}\Bigg(F\Bigg(\frac{\tau_B^{-1}}{\tau_{\phi}^{-1}}\Bigg)-F\Bigg(\frac{\tau_B^{-1}}{\tau_{\phi}^{-1}+2\tau_i^{-1}}\Bigg)-
2F\Bigg(\frac{\tau_B^{-1}}{\tau_{\phi}^{-1}+\tau_i^{-1}+\tau_*^{-1}}\Bigg)\Bigg).
\end{equation}

\noindent Here, $F(z)= lnz+\psi(0.5+z^{-1})$, $\psi(x)$ is the digamma function, $\tau_B^{-1} = 4eDB/\hbar$ [$D$ is the diffusion constant]. The diffusion constant, $D=v_Fl/2$, was determined from the mean free path $l=h/2e^2k_F\rho$, where $\rho$ is the four terminal resistivity at that gate voltage. The corresponding scattering lengths can be calculated using the formula, $L_{\phi,i,*}=\sqrt{D\tau_{\phi,i,*}}$.

\subsection{Temperature dependence of scattering lengths}\label{sec.1}
\begin{figure}[!htbp]
\begin{center}
\includegraphics [width=1\linewidth]{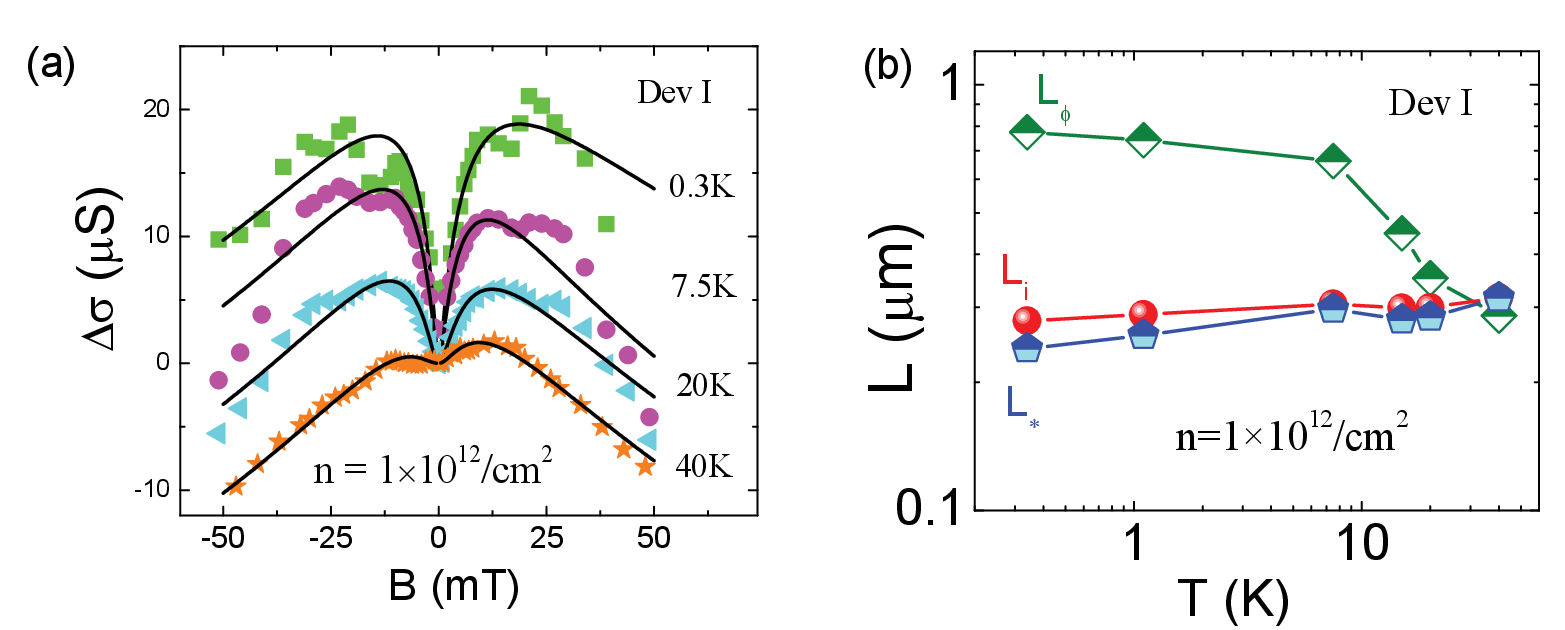}
\end{center}
\end{figure}
\textbf{Fig.~S1.} (a) Magnetoconductivity as a function of temperature taken at an electron density $n = 1\times 10^{12}$/cm$^2$. Solid lines are fit to the weak localization theory (Eq.~\ref{eq1}). (b) The scattering length scales as a function of temperature, extracted from the fits.

Fig.~S1a shows the MC data at a particular density ($1\times 10^{12}$/cm$^2$) for various temperatures for Dev~I and the corresponding fits are shown by solid black lines. With increasing temperature the inelastic phase coherence length ($L_{\phi}$) decreases (Fig.~S1b) and saturates at lower temperatures ($\lesssim 4$~K) at this particular carrier density. However, the intervalley and intravalley scattering lengths remain almost constant with temperature as these are related to the elastic scattering~\cite{Mccann_WL,savchenko_WL}.

\subsection{Gate voltage dependence of $L_{\phi}$ and $L_i$}\label{sec.1}

Fig.~S2a and S2b show the gate voltage/density dependence of MC data at the base temperature of $\sim 10$~mK for Dev~II and Dev~III, respectively. It is clear from both the graphs that with increasing density $L_{\phi}$ increases, whereas, the intervalley scattering length ($L_i$) remains unchanged. The density dependence of the scattering lengths have been studied in detail in Ref.~\cite{inelastic_PRB}. To calculate the fluctuations from a single phase coherent box we have divided the area between the voltage probes in small boxes of area $L_{\phi}^2$, schematically shown in Fig.~S2c. We have used the $L_{\phi}$ value for a particular density, extracted from the fits.

\begin{figure}[!htbp]
\begin{center}
\includegraphics [width=1\linewidth]{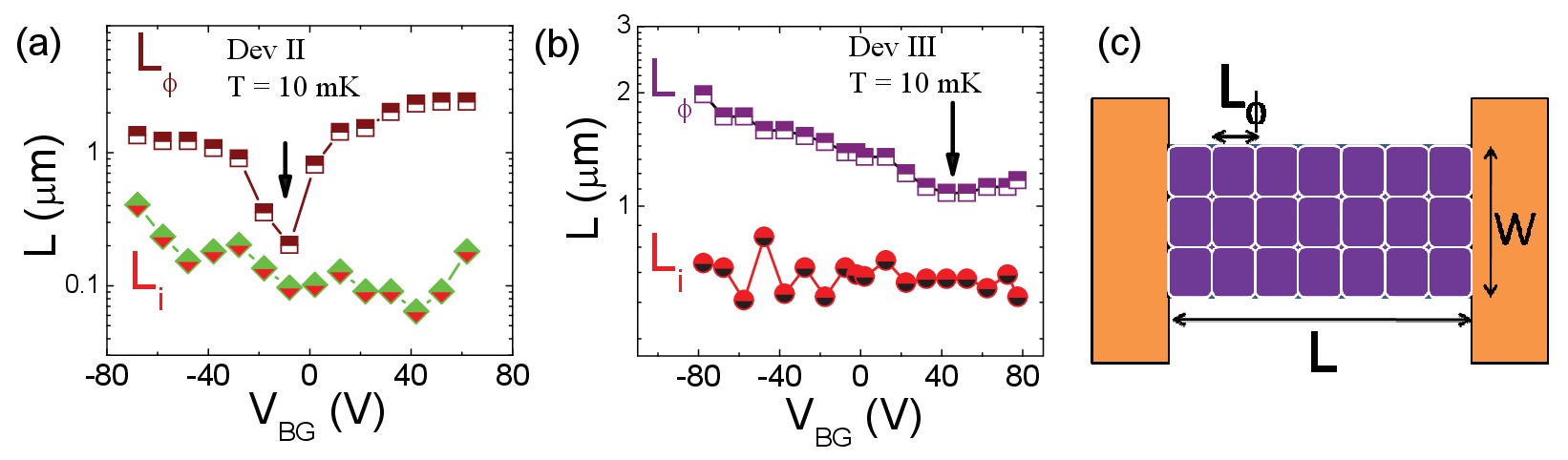}
\end{center}
\end{figure}
\textbf{Fig.~S2.} Gate voltage dependence of $L_{\phi}$ and $L_i$ at $T = 10$~mK, extracted from the weak localization fits, are shown for (a) Dev~II and (b) Dev~III. (c) Schematic describing that the area ($L\times W$) between the voltage probes can be divided into phase coherent boxes of area $L_{\phi}^2$.

\section{Gate Voltage dependence of $\langle \delta G^2 \rangle$}\label{sec.1}

Fig.~S3a and S3b shows the variation of $\langle \delta G^2\rangle$ with gate voltage for Dev~II and Dev~III, respectively. $\langle \delta G^2\rangle$ was calculated using the formula $\langle \delta G^2\rangle = \langle \delta R^2\rangle/\langle R \rangle^4$, where $\langle R \rangle$ is the average four terminal resistance in that gate voltage window. We observe that $\langle \delta G^2\rangle$ is very much device specific and hence, is not a significant parameter to understand the absolute magnitude of UCF.

\begin{figure}[!htbp]
\begin{center}
\includegraphics [width=1\linewidth]{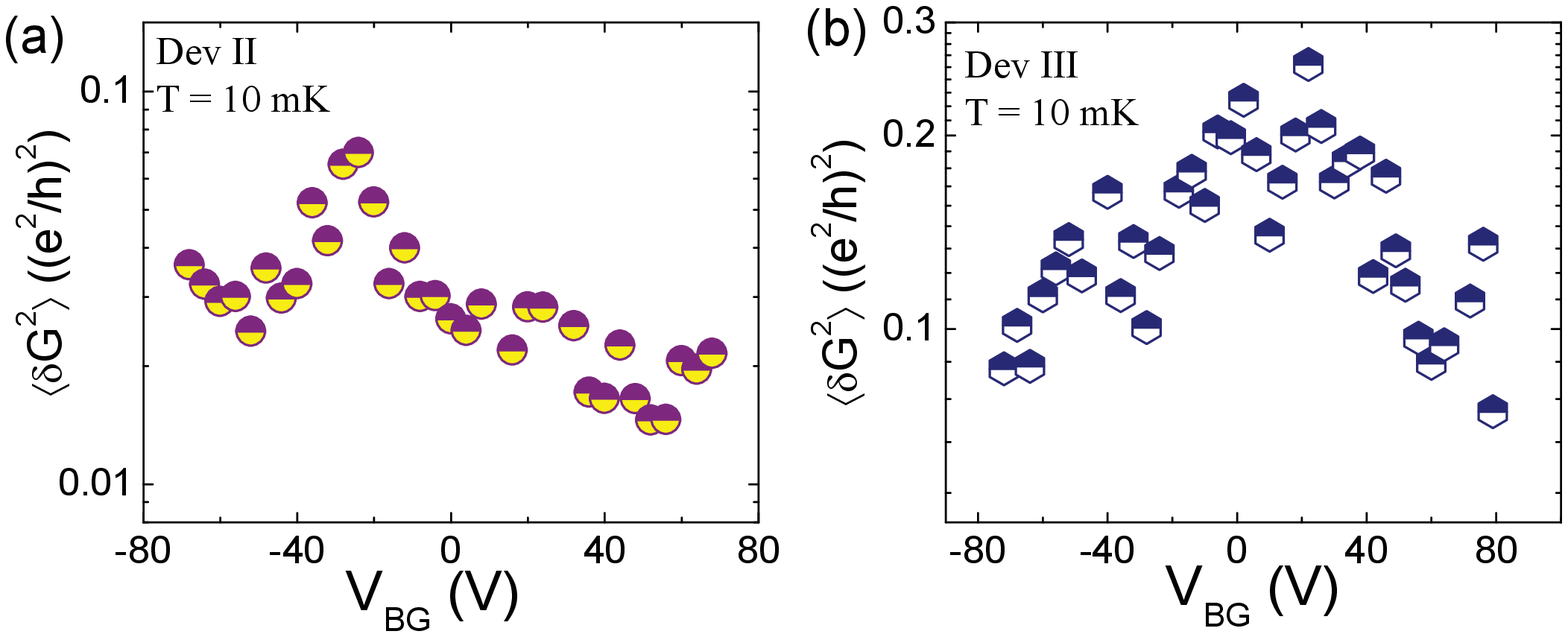}
\end{center}
\end{figure}
\textbf{Fig.~S3.} Variation of $\langle \delta G^2 \rangle$ with gate voltage at $T = 10$~mK, for (a) Dev~II and (b) Dev~III.
